\documentclass[fleqn,usenatbib,revision]{mnras}

\usepackage{graphicx}	
\usepackage{amsmath}	
\usepackage{amssymb}	
\usepackage[dvipsnames]{xcolor}
\usepackage{hyperref}

\usepackage{newtxtext,newtxmath}
\usepackage{ulem}
\usepackage[T1]{fontenc}

\DeclareRobustCommand{\VAN}[3]{#2}
\let\VANthebibliography\thebibliography
\def\thebibliography{\DeclareRobustCommand{\VAN}[3]{##3}\VANthebibliography}

\begin{document}



\title[On the hysteresis effect in transitions]{On the hysteresis effect in transitions between accretion and propeller regimes}

\author[\c{C}\i{}k\i{}nto\u{g}lu \& Ek\c{s}i]{
Sercan \c{C}\i{}k\i{}nto\u{g}lu$^{1}$\thanks{E-mail: cikintoglus@itu.edu.tr},
K. Yavuz Ek\c{s}i$^{1}$\thanks{E-mail: eksi@itu.edu.tr}
\\
$^{1}$Istanbul Technical University, Faculty of Science and Letters,
Physics Engineering Department, 34469, Istanbul, Turkey}

\date{Accepted XXX. Received YYY; in original form ZZZ}

\pubyear{2022}

\label{firstpage}
\pagerange{\pageref{firstpage}--\pageref{lastpage}}
\maketitle

\begin{abstract}
Some observations and numerical simulations of disc-magnetosphere interaction show that accretion can proceed in the propeller regime. When the Alfvén radius is beyond the corotation radius, matter climbs up to the high latitudes where the Alfv\'en surface is inside the equilibrium surface and can accrete. We calculate the fraction of the mass flux in the disc that can accrete onto the neutron star depending on the fastness parameter and the inclination angle between rotation and magnetic axis. We find that, for a narrow range of the fastness parameter, the Alfv\'en and the equilibrium surfaces intersect at two different critical latitudes. 
While the system is transiting from the propeller to the accretion regime (the initial rise of an outburst),
the disc is already thick and the part of the disc between these two critical latitudes cannot accrete.
In transitions from the accretion to the propeller regime (decay of an outburst), the disc is thin, hence,
full accretion of matter proceeds until the Alfv\'en radius moves beyond the equilibrium radius at the disc-midplane.
Therefore, the accretion regime commences at a smaller fastness parameter than it ceases.
As a result, the transition from the propeller to the accretion regime occurs at a luminosity higher than the transition from the accretion to the 
propeller regime. We discuss the implications of our results for spectral transitions exhibited by low-mass X-ray binaries.
\end{abstract}




\section{INTRODUCTION}

Neutron stars in low-mass X-ray binaries accrete matter from a disc \citep{Shakura73,book_Frank02} fed by a low-mass companion \citep{Pringle72}. The interaction of the magnetosphere with the disc modulates the flow of matter onto the magnetic poles allowing for coherent X-ray pulsations revealing the spin frequency of the neutron star to be detected \citep{wijnands98}. These systems are often transients due to thermal-viscous instability within the disc \citep[see e.g.][]{dubus+18} and the X-ray luminosity of the system, determined by the accretion rate onto the neutron star, changes by 4 orders of magnitude during an outburst \citep[see][for reviews]{patruno21,disalvo2022}. As the accretion rate declines, the system is expected to make a transition from the accretion to the propeller stage \citep{Illarionov75,lovelace+99} during which the centrifugal barrier does not allow the matter to fall onto the neutron star.

Axisymmetric (2.5 dimensional) numerical simulations of the propeller regime \citep{Romanova+2004,Ustyugova06,Zanni13,romanova+18} suggest the presence of partial accretion together with the propelling of matter in outflows. This ``partial accretion regime'' is possible since the inner region of the disc becomes thicker and accretion can proceed from the regions away from the midplane. \citet{Menou99} considered reduced accretion due to the propeller effect to address luminosity of neutron star systems at the quiescent stage. \citet{Eksi11} employed this model to address the rapid decline stage in the outburst of SAX~J1808.4--3658. \citet{gungor+17} introduced a `reverse engineering'' method to determine the fraction of mass flux that can accrete onto the star from the  lightcurves of Aql~X--1.  Most recently, \citet{Lipunova+22} presented a detailed discussion of the lightcurves including the effects of irradiation of the disc.

The purpose of the paper is to investigate what fraction of the mass flux in the disc can reach the surface of the neutron star in the fast rotating regime, depending on the rotation rate of the neutron star and to uncover a hysteresis effect in transitions between accretion and propeller regimes. In the next section, the geometric arguments for partial accretion from a spherical flow and a disc is reviewed and improved. In \S~\ref{sec:discuss} we discuss the implications of our results for transient accreting systems with neutron stars.

\begin{figure*}
\centering
\includegraphics{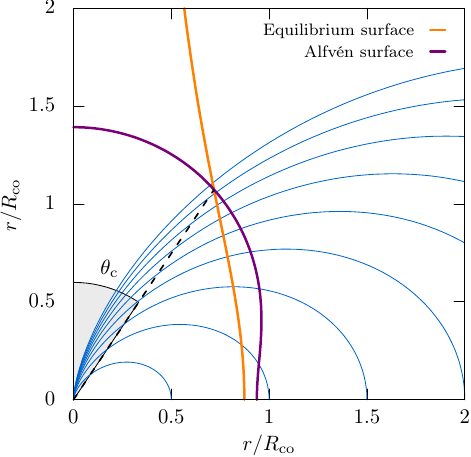}
\includegraphics{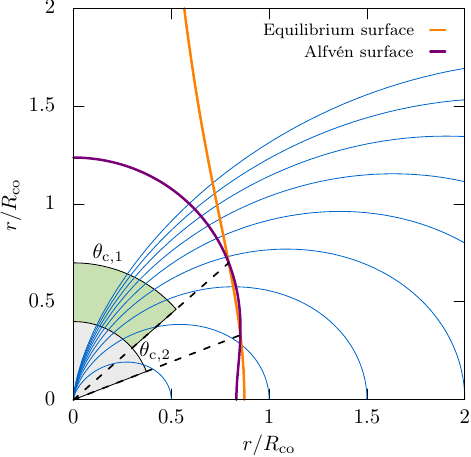}
\caption{
The geometry of the Alfv\'en and the equilibrium surfaces of an aligned rotator for $R_{\rm A}=0.94 R_{\rm co}$ corresponding to $\omega_{\ast}=0.91$ (left panel)
and for $R_{\rm A}=0.83 R_{\rm co}$ corresponding to $\omega_{\ast}=0.76$ (right panel). For $\omega_{\ast}>0.82$ (left panel), the Alfv\'en surface and the equilibrium surface intersect at one co-latitude, $\theta_{\rm c}$, beyond which centrifugal barrier inhibits accretion. In other words, although the Alfv\'en radius is beyond the equilibrium surface at the disc-midplane, a substantial part of the  Alfv\'en surface is inside the equilibrium surface, and accretion can proceed from this region ($\theta<\theta_{\rm c}$). The Alfvén surface can intersect with the equilibrium surface at two points for $0.74 < \omega_{\ast}<0.82$. The right panel shows a sample case where the Alfvén surface is inside the equilibrium surface at two co-latitude ranges ($0-\theta_{\rm c,1}$ and $\theta_{\rm c,2}-\pi/2$). 
}
\label{fig:surfaces}
\end{figure*}

\section{ACCRETION IN THE PROPELLER REGIME}

In this section we derive, from geometrical arguments, the fraction of accretion rate that can reach the surface of the neutron star depending on the fastness parameter, $\omega_* \equiv \Omega_*/\Omega_{\rm K}$ (here, $\Omega_*$ is the angular velocity of the star and $\Omega_{\rm K}$ is the Keplerian angular velocity at the inner radius of the accretion flow at the disc-midplane) and the inclination angle between rotation and magnetic axis, $\alpha$.  We start with the aligned case and depart to the inclined rotator configuration afterwards. This allows us to have a benchmark to check the calculations of the inclined rotator case at the $\alpha=0$ limit.

\subsection{Aligned rotator}

We first assume that the magnetic moment of the dipole is aligned with the rotation axis of the star. We further assume, for simplicity, that the presence of the disc does not change the field configuration from dipole which is obviously an oversimplification and results in a toy model. The only justification is that these also are the assumptions inherent in the derivation of the Alfv\'en radius. 

The magnetic field ${\mathbf B}$ for an aligned dipole can be written as
\begin{equation}
\mathbf{B}=\frac{\mu _{\ast} }{r^3}(2\cos \theta\,\hat{\mathbf{e}}_r
+\sin \theta\,\hat{\mathbf{e}}_{\theta})
\label{eq:B}
\end{equation}
In spherical coordinates ${\rm d}\mathbf{r}={\rm d}r\,\hat{\mathbf{e}}_{r}+r {\rm d}\theta\, 
\hat{\mathbf{e}}_{\theta }+r\sin \theta{\rm d}\phi\,\hat{\mathbf{ e}}_{\phi }$ the field lines are described by
\begin{equation}
\frac{{\rm d}r}{B_{r}}=\frac{r\, {\rm d}\theta }{B_{\theta }}=\frac{r\sin \theta \, {\rm d}\phi }{%
B_{\phi }}\,.
\end{equation}
This can be integrated to give
\begin{equation}
r=C\sin^2 \theta
\label{eq:dipole}
\end{equation}
where $C$ labels different field lines. 
The magnitude of the poloidal magnetic field is
\begin{equation}
|B_{\rm p}|^2 = B_r^2+B_{\theta}^2 
=\frac{\mu_{\ast }^2} {r^6}\left( 1 + 3\cos^2 \theta \right)
\end{equation}
where we used equation~\eqref{eq:B}.
The inner radius of a thin disc can be determined by the condition
of the material and magnetic stresses at the disc-midplane \citep[see equation 42 in][]{gho79a},
\begin{equation}
       \frac{1}{4\pi r^2}\dot{M} r^2\Omega_K  = \frac{1}{4\pi}B_\phi B_\theta \Delta r\,,    
\end{equation}
Here, $\Delta r$ is the width of the transition region where
the disc flow deviates from the Keplerian motion.
For a non-thin disc, we generalise this condition as
\begin{equation}
       \frac{1}{4\pi r^2}\dot{M} r^2\Omega_K  = \frac{1}{4\pi}\gamma |B_p|^2 \Delta r\,,    
\end{equation}    
where
\begin{equation}
   \gamma \equiv \frac{B_\phi}{\sqrt{B_r^2+B_{\theta}^2}}
\end{equation}
is the ratio of the toroidal magnetic field to the strength of the poloidal magnetic field.
This can be used to define an Alfv\'{e}n surface
\begin{equation}
r_{\mathrm A} = R_{\rm A}\left(1+3\cos^2 \theta \right)^{2/7}    
\label{eq:r_A}
\end{equation}
where
\begin{equation}
R_{\rm A}=\xi \left( \frac{\mu _{\ast }^{2}}{\sqrt{GM_{\ast }}\dot{M}}\right)^{2/7}.
\end{equation}
is the Alfv\'en radius, the radius of the Alfv\'en surface at the disc-midplane and $\xi\equiv \gamma\Delta r/r$ is a numerical factor at the order of unity.
We will consider $\xi$ as unity throughout the paper. 
\begin{figure}
\centering
\includegraphics{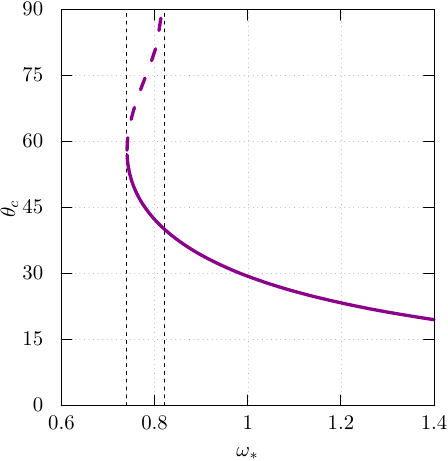}
\caption{
The dependence of the critical angle $\theta_{\rm c}$ on the fastness parameter, $\omega_{\ast}$, for the aligned rotator. For  $0.74 < \omega_{\ast}<0.82$
(the interval between vertical black dashed lines), the critical angle $\theta_{\rm c}$ is doubly-valued as Alfv\'en surface cuts the equilibrium surface at two locations. The dashed purple line represents the larger root of $\theta_c$ for a given $\omega_{\ast}$. For $\omega_{\ast}=0.74$ it is seen that  $\theta_{\rm c} \simeq 58^\circ$.}
\label{fig:thetac}
\end{figure}

\begin{figure*}
\centering
\includegraphics{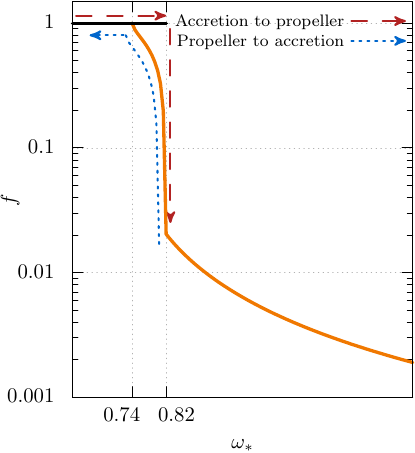}
\includegraphics{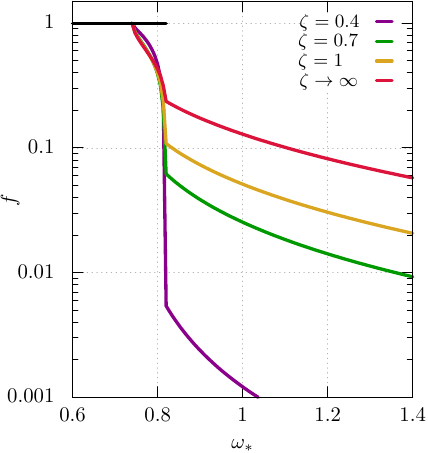}
\caption{The fraction of mass flux that can accrete onto the star depending on $\omega_{\ast}$
for the aligned rotator.
The black solid line represents full accretion, i.e., $f=1$, for all cases.
In the left panel, the hysteresis effect is demonstrated on a representative example (for $\zeta=0.5$). 
When the system transits from the accretion to the propeller regime, the fraction values follow the red arrows. 
On the other hand, the fraction values follow the blue arrows when the system transits from the propeller to the accretion regime. The right panel shows the fractions for different values of $\zeta$. 
}
\label{fig:fraction}
\end{figure*}

For a point mass rotating with the stellar angular velocity, $\Omega_\ast$, 
the acceleration towards the rotation axis is 
$\mathbf{a}_{\rm c}=\Omega _{\ast}^{2} r\sin \theta 
\left(\sin\theta\,\hat{\mathbf{e}}_r+\cos\theta\,\hat{\mathbf{e}}_\theta\right)
$.
The balance of this acceleration with the gravitational acceleration,
$\mathbf{a}_g =-\left(GM/r^2\right)\hat{\mathbf{e}}_r$,
along the magnetic field
defines the equilibrium surface,
\begin{equation}
r_{\rm eq} = (2/3)^{1/3} R_{\rm co} \sin^{-2/3}\theta\,,
\label{eq:r_c}
\end{equation}
\citep{Lyutikov22} where 
the Keplerian angular velocity of the disc
matches the stellar angular velocity,
the so-called corotation radius, is given by
\begin{equation}
R_{\rm co} =\left( \frac{GM_{\ast}}{\Omega_{\ast}^2} \right)^{1/3}.
\end{equation}
Note that the above condition for the equilibrium surface is not valid at the disc-midplane where the magnetic field
is perpendicular to the gravitational and centrifugal accelerations. However, equation~\eqref{eq:r_c} can be used at the disc-midplane by the continuity.
Thus, the equilibrium radius at the disc-midplane is $R_{\rm eq} =(2/3)^{1/3} R_{\rm co}$.
Accordingly, the propeller regime starts when $R_{\rm A}$ is larger than $(2/3)^{1/3} R_{\rm co}$ 
rather than $R_{\rm co}$.

The intersection of the equilibrium surface with the Alfv\'en surface defines a critical angle $\theta_{\rm c}$ below which the disc material does not meet with a equilibrium surface and so can accrete onto the star (see Fig.~\ref{fig:surfaces}). This critical angle depends on the fastness parameter given implicitly by
\begin{equation}
\omega_{\ast}^{-1} =\sqrt{\frac32} \sin \theta_{\rm c} \left( 1 + 3 \cos^2 \theta_{\rm c} \right)^{3/7},
\label{eq:wt}
\end{equation}
which is found by $r_{\rm eq}=r_{\rm A}$
given in equations \eqref{eq:r_A} and \eqref{eq:r_c}, and referring the definition of the fastness parameter%
\footnote{Note that \citet{Menou99} assumes $r_{\rm A} \simeq R_{\rm A}$ (spherical magnetosphere) and obtains $\omega_{\ast}^{-1} = \sin \theta_{\rm c}$ analytically which is accurate only for $\theta_{\rm c} \simeq \pi/2$.
}. 
The numerical solution of $\theta_{\rm c}$ from this equation is shown in Fig.~\ref{fig:thetac}. 
Note that, for $0.74 < \omega_{\ast}<0.82$, this equation has two solutions for $\theta_{\rm c}$ since the Alfv\'en surface intersects the equilibrium surface at two distinct altitudes. 
The matter within these two critical altitudes is beyond the equilibrium surface and hence is expelled.
Accordingly, the matter can accrete from two distinct regions;
between the disc-midplane and the higher altitude, and between the lower altitude and the spin axis (see the right panel of Fig.~\ref{fig:surfaces}).
We assume the disc flow might be channelled onto the star from these two regions simultaneously.
On the other hand, when $\omega_{\ast}>0.82$,
the Alfv\'en radius is beyond the equilibrium radius at the disc-midplane
and the Alfv\'en surface intersects with the equilibrium surface at one altitude,
therefore, only the matter between $0-\theta_{\rm c}$ can accrete onto the star
(see the left panel of Fig.~\ref{fig:surfaces}).

The fraction of mass inflow that can accrete onto the star is then given by
\begin{align}
f  \equiv \frac{\dot{M}_{\ast}}{\dot{M}} =&\,
 \frac{2\int_0^{\theta_{\rm c,1}} 2\pi r^2 \sin \theta \rho(r,\theta) v(r,\theta)\,{\rm d} \theta}
 {2\int_0^{\pi/2} 2\pi r^2 \sin \theta \rho(r,\theta) v(r,\theta)\,{\rm d} \theta}
 \notag \\
& +\frac{2\int_{\theta_{\rm c,2}}^{\pi/2} 2\pi r^2 \sin \theta \rho(r,\theta) v(r,\theta)\,{\rm d} \theta}
 {2\int_0^{\pi/2} 2\pi r^2 \sin \theta \rho(r,\theta) v(r,\theta)\,{\rm d} \theta}
\label{eq:fract} 
\end{align}
\citep{Menou99} where
$\theta_{\rm c,1}$ and $\theta_{\rm c,2}$ are roots of equation~\eqref{eq:wt}, $\rho$ is the density, $v$ is the radial velocity and the factor 2 stands for the possibility of accretion onto both poles. 
When $\omega_*=0.82$, the higher altitude solution, $\theta_{\rm c,2}$, goes to $\pi/2$ and vanishes for larger fastness parameter values. Thus, the second term in equation~\eqref{eq:fract} vanishes for $\omega_*\geq 0.82$. When $\omega_*=0.74$, two roots become identical and the fraction becomes $1$, therefore, the fraction
is always unity for $\omega_*\leq 0.74$.

The partial accretion of a spherical flow onto a rapidly rotating neutron star was first calculated by \citet{Lipunov76}.
For disc accretion, we follow \citet{Menou99} and assume $v(r,\theta)=v_{\rm c}(r)\sin^2 \theta$. Additionally, we use the isothermal vertical disc solution $\rho(r,\theta)= \rho_{\rm c}(r)\,\exp(-r^2 \cos^2\theta/H^2)$ \citep{book_Frank02}  where $H$ is the thickness of the disc. 
By plugging these in equation~\eqref{eq:fract} and evaluating integrals 
at $r=r_{\rm c}$ where the matter is channeled from the disc
we obtain
\begin{align}
f=&\,
1-
\frac{
\sqrt{\pi}\left[1-2/\zeta^{2}\right]\mathrm{erf}\left(\cos\theta_{\rm c,1}/\zeta \right)
-2\zeta^{-1}\cos\theta_{\rm c,1}\,{\rm e}^{-\cos^2\theta_{\rm c,1}/\zeta^{2}}}
{
\sqrt{\pi}\left[1-2/\zeta^{2}\right]\mathrm{erf}\left(\zeta^{-1}\right)
-2\zeta^{-1} {\rm e}^{-1/\zeta^{2}}
}
\notag \\
&+\frac{
\sqrt{\pi}\left[1-2/\zeta^{2}\right]\mathrm{erf}\left(\cos\theta_{\rm c,2}/\zeta \right)
-2\zeta^{-1}\cos\theta_{\rm c,2}\,{\rm e}^{-\cos^2\theta_{\rm c,2}/\zeta^{2}}}
{
\sqrt{\pi}\left[1-2/\zeta^{2}\right]\mathrm{erf}\left(\zeta^{-1}\right)
-2\zeta^{-1} {\rm e}^{-1/\zeta^{2}}
}
\,,
\label{eq:fdisc}
\end{align}
for disc accretion where $\zeta\equiv H(r_{\rm c})/r_{\rm c}$ is the thickness parameter of the disc
at $r=r_{\rm c}$ and $\mathrm{erf}$ is the error function,
\begin{equation}
\mathrm{erf}(x)=\frac{2}{\pi}\int^{x}_0 {\rm e}^{-t^2}\,\mathrm{d}t.
\end{equation}
In the limit of $\zeta\rightarrow\infty$ (spherical accretion) the fraction of mass inflow becomes
\begin{equation}
f = 1-\cos\theta_{\rm c,1}+\cos\theta_{\rm c,2}\,,  
\end{equation}
while, in the limit of thin disc, i.e.,~$\zeta\rightarrow 0$, 
\begin{equation}
f=\begin{cases}
1,\quad &\omega_*<0.82,\\
0,\quad &\omega_*>0.82.
\end{cases}
\end{equation}

The system is in full accretion regime ($f=1$) when $\omega_*<0.74$ since the Alfv\'en surface is 
fully inside the equilibrium surface. When $\omega_*>0.82$, the Alfv\'en surface coincides with the equilibrium surface 
at one latitude (see the left panel of Fig.~\ref{fig:surfaces}). To commence partial accretion, the disc has to fill the latitudes until $\theta_{\rm c,1}$.
We assume that the disc is thin ($\zeta\rightarrow 0$) in the full accretion regime. However, when the Alfv\'en radius
moves beyond the equilibrium surface ($\omega_*>0.82$), the matter cannot accrete onto the star, therefore, the disc would get thicker due to the accumulated matter in the viscous time-scale
which is much shorter than the time-scale of observations. As a result, the disc might fill latitudes above $\theta_c$, hence,
partial accretion might become possible.
Therefore, if the inner part of the disc is fully inside the equilibrium surface or it moves outward from there, we assume that 
the disc is thin ($\zeta\rightarrow0$) until $\omega_*=0.82$.
If the Alfv\'en radius is beyond the equilibrium surface or it moves inward from there, we assume the disc has a non-negligible thickness until $\omega_*=0.74$.
This model can be summarised as
\begin{equation}
    \begin{aligned}
    \zeta\rightarrow 0,\quad &\text{if}\;\,\omega_*<0.74,\\
    \zeta\rightarrow 0,\quad &\text{if}\;\,0.74<\omega_*<0.82\,\text{ and $\omega_*$ ascending from $0.74$},\\
    \zeta=\text{Finite},\quad &\text{if}\;\,0.82<\omega_*,\\
    \zeta=\text{Finite},\quad &\text{if}\;\,0.74<\omega_*<0.82\text{ and $\omega_*$  descending from $0.82$.}
\end{aligned}
\end{equation}
We calculate $f$ accordingly and report its dependence onto the fastness parameter in Fig.~\ref{fig:fraction}.
While the system transits from the propeller regime to the accretion regime
and during the propeller,
we assume the disc has a finite constant thickness otherwise it is thin.
As a result of the different thickness parameter of the disc in transitions,
the fraction is doubly-valued for $0.74<\omega_*<0.82$ and takes different values depending on whether $\omega_*$ is ascending or descending.
Moreover, the partial accretion requires the inner part of the disc to be thick such as $\zeta \gtrsim 0.4$.

\begin{figure}
\centering
\includegraphics[width=0.48\textwidth]{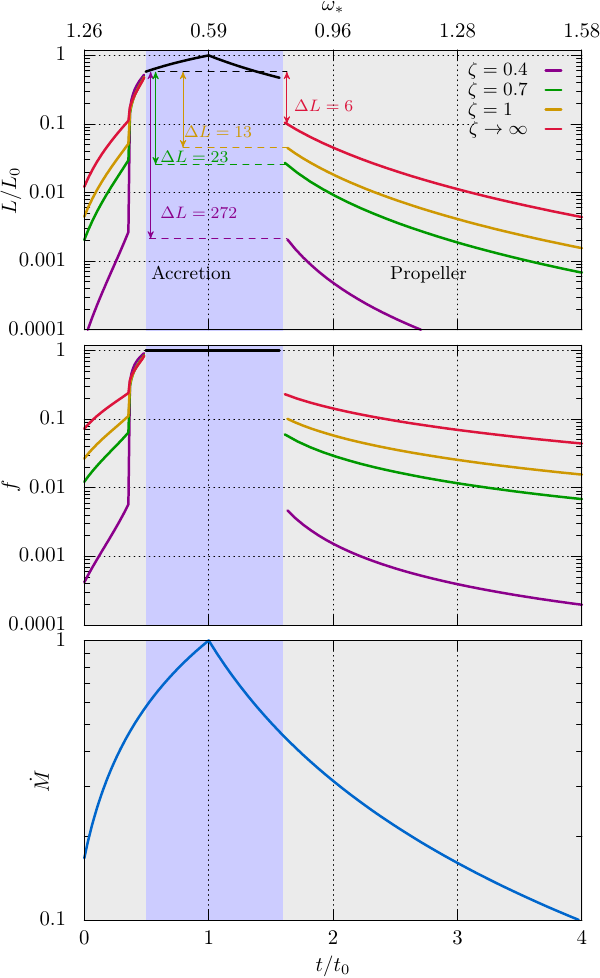}
\caption{
Top panel: The evolution of the normalized lightcurve for aligned rotators. The black solid line represents the full accretion stage, i.e., $f=1$, for all cases.
Here, $\Delta L$ is the ratio of the luminosities of transition from the propeller to accretion and the transition from accretion to the propeller regimes.
Middle panel: The evolution of the fraction for the same cases.
Bottom panel: The evolution of the mass accretion rate.
}
\label{fig:lightcurve}
\end{figure}

In Fig.~\ref{fig:lightcurve}, we report the evolution of the lightcurves to
demonstrate the effect of the partial accretion.
The luminosity produced by the accretion is given as
\begin{equation}
L=\frac{GM_*\dot{M}}{R_*}f
\end{equation}
where $G$ is the gravitational constant, $M_*$ and $R_*$ are, respectively, the mass and the radius of the star.
We assume the mass accretion rate evolves as
\begin{align}
\dot{M}=
\begin{cases}
\displaystyle \frac{\dot{M}_0}{6}\left(1+\frac{5t}{t_0}\right), 
\quad & t<t_0\,, \\
\displaystyle \dot{M}_0\left(\frac{t}{t_0}\right)^{-5/3}, 
\quad & t>t_0\,,
\end{cases}
\end{align}
where $\dot{M}_0$ and $t_0$ are some arbitrary constants.
This toy model ensures that the mass accretion rate linearly increases until $t=t_0$, then,
it decreases rapidly (see the bottom panel of Fig.~\ref{fig:lightcurve}). 
So, we can observe the transition from the propeller regime to the accretion regime
and then to the propeller regime again.
Additionally, we set the inner radius of the accretion flow
to $0.7R_{\rm co}$ at $t=t_0$ in all cases.

As mentioned above, the fraction of the mass inflow that can accrete onto the star takes different values
in transitions ($0.74<\omega_*<0.82$).
Therefore, the luminosity produced by the accretion would be different between the transition from the accretion to the propeller regime and the vice-versa transition as the full accretion commence or ceases at different fastness parameters (see the middle panel of Fig.~\ref{fig:lightcurve}).
The ratio of the luminosities of these two transitions is 6 for $\zeta\rightarrow \infty$ limit, and it increases as the thickness parameter of the disc reduces.
Other than the thickness parameter of the disc, our results for the luminosity ratios
depend on the choice of the inner radius of the accreting flow at $t=t_0$ while 
they are not affected by the toy model of the mass-accretion rate we employed.
The difference between the transition luminosities exposes a hysteresis effect in the lightcurve as can be seen from Fig.~\ref{fig:lightcurve}.

\subsection{Inclined rotator}
The dipole magnetic field of the star
can be written in a general form
as
\begin{equation}
\mathbf{B}=\frac{\mu}{r^3}\left[ 3\left(\hat{{\boldsymbol\mu}}\cdot\hat{\mathbf{e}}_r\right)\hat{\mathbf{e}}_r-\hat{{\boldsymbol\mu}} \right],
\end{equation}
where $\hat{{\boldsymbol\mu}}$ is the unit magnetic dipole moment vector.
If we choose the coordinate system such that
$z$-axis is the rotation axis of the spherical star,
the unit magnetic dipole moment vector can be written as
\begin{equation}
\hat{{\boldsymbol\mu}}=\sin\alpha\cos\chi\,\hat{\mathbf{e}}_x
                      +\sin\alpha\sin\chi\,\hat{\mathbf{e}}_y
                      +\cos\alpha\,\hat{\mathbf{e}}_{z}\,,
\end{equation}
where $\alpha$ is the inclination angle between the rotation axis and the magnetic dipole moment,
and $\chi=\Omega t$ is the angle between the $x-z$ plane and the magnetic dipole moment.

\begin{figure}
\centering
\includegraphics{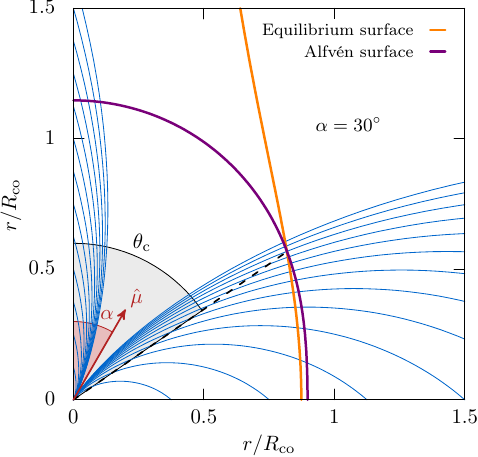}
\caption{
The geometry of the averaged Alfv\'en and the averaged equilibrium surfaces of an inclined rotator for $\omega_{\ast}=0.85$.
}
\label{fig:surfaces_inclined}
\end{figure}

\begin{figure}
\centering
\includegraphics{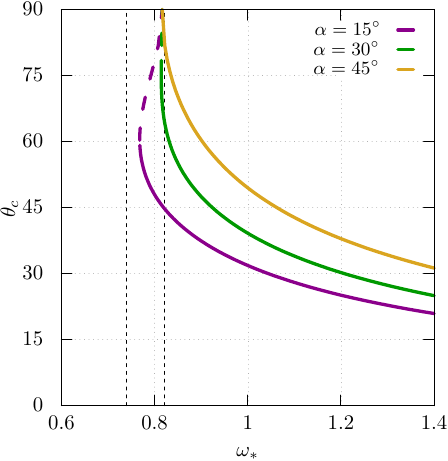}
\caption{
The dependence of the critical angle $\theta_{\rm c}$ on the fastness parameter, $\omega_{\ast}$, for the inclined rotator.
Dashed lines represent the larger root of $\theta_c$ for a given $\omega_{\ast}$ for each alignment angle.
Vertical dashed lines bound the interval of $0.74 < \omega_{\ast}<0.82$.}
\label{fig:thetac_inclined}
\end{figure}

Since the timescale of observations of lightcurves are much longer than the spin period of the star, we calculate the time-averaged Alfv\'en surface over a period,
\begin{align}
\left\langle r_{\rm A}\right\rangle
=\,&R_{\rm A}\left[ 
\frac{5}{2}-\frac{3}{2}\cos^2\alpha
-\frac{3}{2}\cos^2\theta\left(1-3\cos^2\alpha\right)
\right]^{2/7}\,,
\end{align}
as well as the equilibrium surface,
\begin{align}
\left\langle r_{\rm eq}\right\rangle
=&\,(2/3)^{1/3} R_{\rm co} \sin^{-2/3}\theta\,.
\end{align}
Note that the average size of the magnetosphere at the disc-midplane is
$R_{\rm A}\left(\frac{5}{2}-\frac{3}{2}\cos^2\alpha
\right)^{2/7}$ for the inclined rotator rather than the $R_{\rm A}$,
hence, $\omega_*=\left[\left(\frac{5}{2}-\frac{3}{2}\cos^2\alpha
\right)^{2/7} R_{\rm A}/R_{\rm co}\right]^{3/2}$. Hence the critical fastness parameter for transition to the propeller stage is $\omega_{\rm c}=\sqrt{2/3}\simeq 0.82$.
Accordingly, the averaged Alfv\'en and the averaged equilibrium surfaces intersect (see Fig.~\ref{fig:surfaces_inclined}) at
\begin{align}
\omega_*^{-1}
=&\,
\sqrt{\frac32} 
\sin\theta_{\rm c}
\left[ 
\frac{5}{2}-\frac{3}{2}\cos^2\alpha
\right]^{-3/7}
\notag \\
&\times\left[ 
\frac{5}{2}-\frac{3}{2}\cos^2\alpha
-\frac{3}{2}\cos^2\theta_{\rm c}\left(1-3\cos^2\alpha\right)
\right]^{3/7}\,,
\end{align}
\citep[see also][]{Abolmasov2020}.
As a result of the time average,
dependency on the $\chi$ angle vanishes. Furthermore,
the intersection of the averaged Alfv\'en and the averaged equilibrium surfaces
are symmetric for $\theta\rightarrow \pi-\theta$
which is different than \citet{Lyutikov22}.
We find that the critical angle $\theta_{\rm c}$ is doubly-valued for a narrower range 
of the fastness parameter as the inclination angle increases (see Fig.~\ref{fig:thetac_inclined}).
Moreover, the critical angle is not doubly-valued for $\alpha\gtrsim 30^\circ$ implying that the hysteresis effect would not occur in highly inclined rotators.

\begin{figure*}
    \centering
    \includegraphics{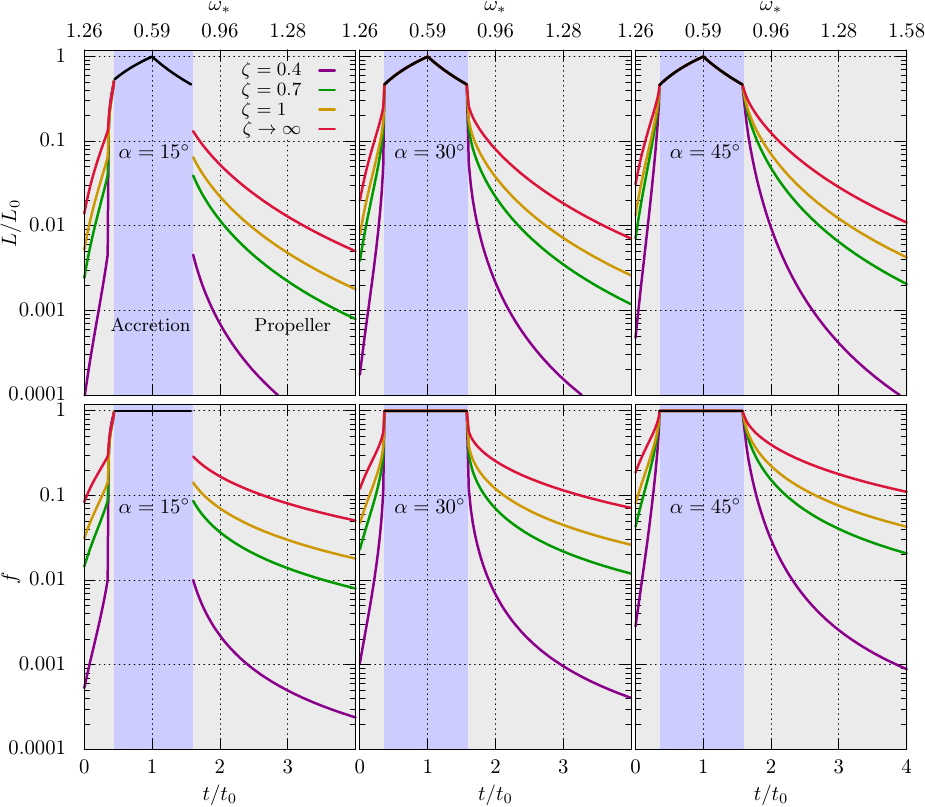}
    \caption{Top panel: The evolution of the normalized lightcurve of inclined rotators.
    The black solid line represents full accretion stage, i.e., $f=1$, for all cases.
    Bottom panel: The evolution of the fraction for the same cases.
    \label{fig:inclined_lightcurve}}
\end{figure*}

The fraction of mass inflow that can accrete onto the star is given by the 
same expressions, equation~\eqref{eq:fdisc}, as in the aligned rotator.
Figure \ref{fig:inclined_lightcurve} reports the evolution of the lightcurve in the case of the inclined magnetic field for different inclination angles and disc thickness values.
We use the same toy model for the mass-accretion rate as in the previous section and set the inner radius of the accretion flow
to 
$\frac{7}{10}\left(\frac{5}{2}-\frac{3}{2}\cos^2\alpha
\right)^{-2/7}R_{\rm A}$ at $t=t_0$.
The difference between transition luminosities decreases
as the inclination angle increases. 
For instance, the ratio of the transition luminosities
is $4$ for $\zeta\rightarrow\infty$ and it is 
$118$ for $\zeta=0.4$ when $\alpha=15^\circ$.
When $\alpha>30^\circ$, the Alfv\'en surface does not intersect the equilibrium surface at two latitudes for any values of the fastness parameter as mentioned above. Therefore, the fraction is no longer doubly-valued and the hysteresis effect is lost.

\section{DISCUSSION} 
\label{sec:discuss}

We have given geometrical arguments to calculate the fraction $f$ of mass-flux in the disc that can reach the surface of the neutron star depending on the fastness parameter, $\omega_*$ and inclination angle $\alpha$. We have seen that, for a range of the fastness parameter near the transition region ($\omega_* \simeq 0.82$ in the aligned case), the fraction $f$ is two-valued since the thickness of the disc is different in transitions and the Alfv\'en surface intersects the equilibrium surface at two different critical angles, $\theta_{\rm c}$. If the system transits from the propeller to the accretion regime, such as at the commence of an outburst, the disc is already thick and some part of the disc might be outside the equilibrium surface. If, on the other hand, the system is transiting from the accretion to the propeller stage, such as at the decay of an outburst, the disc is thin and fully inside the equilibrium surface. This leads to a hysteresis effect in the lightcurves of transient accreting systems in the sense that the transition from the propeller to the accretion regime occurs at a luminosity higher than the luminosity at which the transition from the accretion to the propeller regime occurs. 

Our results show that the thicker is the inner disc, the smaller is the luminosity difference in the transitions (see Fig.~\ref{fig:lightcurve}). The luminosity difference also decreases with the inclination angle (see Fig.~\ref{fig:inclined_lightcurve}) vanishing at $\alpha\simeq 30^\circ$ where the hysteresis effect is lost since the critical angle at which the averaged Alfvén surface intersects with the averaged equilibrium surface is not doubly-valued for such large values of $\alpha$ (see Fig.~\ref{fig:thetac_inclined}).

We would like to note that our results depend on several assumptions, e.g.\ the assumption about the shape of the magnetosphere as an ideal dipole is expected to be modified in the presence of the disc material \citep[see][]{Lyutikov22}. 
Moreover, 
by crudely generalising the condition of balance of stresses given in equation~(42) of \citet{gho79a},
we obtain expression \eqref{eq:r_A} and use it
to determine the Alfv\'en surface for an accretion disc which is not necessarily thin
although the Alfv\'en surface of a non-thin disc would require more detailed investigations.

Also, we use arbitrary constant thickness parameter values to determine the fraction of the mass-accretion rate.
However, the accumulation of the matter in the propeller regime causes the disc to get thicker.
Therefore,
the fraction of the mass-accretion rate, as a boundary condition of the mass-loss
determines the dynamics of the disc. Beside the thickness parameter,
the simplistic functions we employed for the density and the velocity of the flow may in turn different.
Moreover, the mass-accretion rate might be effected by the change of the thickness of the disc.
Hence our solution for $f$ and the resulting lightcurves may be different in a self-consistent solution. As a follow up to our general relativistic magnetohydrodynamics (GRMHD) simulations \citep{Cikintoglu+22} with \texttt{Black Hole Accretion Code} (BHAC) \citep{Porth17}, we will extend our analysis to rotating neutron stars and study the partial accretion regime. 

We stress however that the presence of the hysteresis effect in transitions between the accretion and the propeller regimes does not depend on the details of the disc flow and on our specific assumptions about the $\theta$ dependence of $v$ or $\rho$, but on
the thickness parameter being different in transitions, and
on the shape of the magnetosphere, as given in equation \eqref{eq:r_A}, being non-spherical allowing for the intersection of the equilibrium surface at two different altitudes (consider the peanut shape the magnetosphere would have if Fig.~\ref{fig:surfaces} is drawn in 3-dimensions). If a spherical magnetosphere ($r_A \simeq R_A$) is employed for the sake of simplicity, the hysteresis effect is lost.

The most trivial prediction of the model presented here is that, for systems where $\alpha \lesssim 30^\circ$, the transition to the propeller regime would be exhibited in a lightcurve with an abrupt drop to a lower luminosity followed by a slower decay. 

It is tempting to associate the hysteresis effect we discuss in this work with the hysteresis effect observed by \citet{maccarone03} in the spectral transitions of Aql~X--1. Since then many systems exhibited the hysteresis effect the most recent being 4U~1730--22 \citet{Chen22}. These systems exhibit transitions from the low-hot state, associated with the propeller stage, to the high-cold state associated with the accretion stage. These works report that the former transitions occur at a few times higher luminosity compared to the latter transitions, a hysteresis effect that is akin to what we present here.

Although compelling, we must be cautious in making the above association since such hysteresis effect in spectral transitions is observed also from systems where the accreting objects are black holes \citep[see e.g.][]{munoz+14}. Although black holes can not have magnetic fields themselves (no-hair theorem), they can have magnetospheres \citep{blandford77,komissarov04,crinquand+20,bransgrove+21,crinquand22} coupled to the inner disc, but it is unlikely that these magnetospheres would allow black holes to experience a propeller stage similar to the neutron stars. Thus the hysteresis effect for transitions between accretion and propeller regimes that we propose here can not be a favourable explanation of the hysteresis effect in spectral transitions observed in LMXBs if one insists on a common mechanism working both for black hole and neutron star accretors.

\section*{Acknowledgements}
We acknowledge support from the Scientific and Technological Research Council of Turkey (TÜBİTAK) with project number 112T105. We thank Luciano Rezzolla for a careful reading of an early version of the manuscript. 
\section*{Data availability}

This is a theoretical paper that does not involve any new data. The
model data presented in this article are all reproducible.


\bibliographystyle{mnras}
\bibliography{refs.bib} 



\bsp	
\label{lastpage}
\end{document}